\documentstyle[aps]{revtex}
\begin{document}
\draft
\title{
Superconductor-Metal-Insulator Crossover 
in Disordered Thin Films
\footnote{This paper has been submitted to 
the Short Note section of J. Phys. Soc. Jpn.}
}
\author{
O. Narikiyo
}
\address{
Department of Physics, 
Kyushu University, 
Fukuoka 810-8560, 
Japan
}

\vskip 10pt

\date{
Oct. 4, 2005
}

\vskip 10pt

\maketitle
%----------------------------------------------------------------
\begin{abstract}
Superconductor-Metal-Insulator crossover 
in disordered thin films 
is discussed on the basis of a random 
junction-network model. 
\end{abstract}
%----------------------------------------------------------------
\vskip 30pt

The superconductor-insulator (S-I) transition in thin films 
is one of the typical phenomena of quantum phase transition.~\cite{RMP} 
Recently a metallic phase 
intervening between superconducting and insulating phases 
has been drawing attention.~\cite{PD} 
Although several theories~\cite{PD} 
modifying the idealized theory~\cite{RMP} for the S-I transition 
have been proposed, 
present understanding of the metallic phase 
is far from conclusive. 
In this Short Note we also propose a simple phenomenological model 
and try to understand the temperature dependence of the resistance 
from which we judge the phase at zero temperature, 
superconductor (S) or metal (M) or insulator (I). 

Our phenomenological model is a combination of two models. 
One is the random Josephson-junction-network model~\cite{Dynes} 
for disordered superconductors 
and the other is the random resister-network model~\cite{Meir} 
for disordered metals or insulators. 
We consider a network 
consisting of superconducting islands 
linked by two types of junctions, J (Josephson) and N (normal). 
At the type-J junction 
the current is carried by Cooper pairs, 
while at the type-N junction 
by normal electrons. 
The type-N junction in our model is expected 
for the case where the distance between islands is so large 
that the Cooper pairs are broken in the junction. 
In real materials 
the type-N junction is also expected 
for the case where the size of the island is so small 
that the Cooper pairs are not formed there. 
In our model 
only the resistance caused by the junctions 
is taken into account 
and the resistance drop 
at around the superconducting transition temperature $T_{\rm c}$ 
of the islands is neglected. 
In the following we set $\hbar = k_{\rm B} = 1$. 

At a type-J junction~\cite{Dynes} 
the resistance $R(T)$ as a function of the temperature $T$ 
is assumed to be given by the Ambegaokar-Halperin formula 
\begin{equation}
R(T) = R_{\rm N} / [ I_0(E_{\rm J}(T)/T) ]^2, 
\end{equation}
where $I_0(x)$ is the modified Bessel function of order $0$. 
The Josephson-coupling energy $E_{\rm J}(T)$ 
is given by the Ambegaokar-Baratoff form 
\begin{equation}
E_{\rm J}(T) = {1 \over 4} {R_0 \over R_{\rm N}} 
               \Delta(T) \tanh{\Delta(T) \over 2T}, 
\end{equation}
where $R_0$ is the quantum unit of the resistance, 
$R_0 = \pi / e^2$, 
and $\Delta(T)$ is the BCS gap function. 
This model is appropriate for relatively thick films 
where contacts among islands are relatively strong. 
Since the normal resistance $R_{\rm N}$ between islands 
saturates at low temperatures 
and exhibits only weak temperature dependence 
for such a strong contact, 
we neglect the temperature dependence of $R_{\rm N}$ 
in accordance with the former study.~\cite{Dynes} 
In contrast to this good metallic behavior 
a poor metallic junction disscussed below 
exhibits some temperature dependence. 

At a type-N junction~\cite{Meir} 
the resistance is assumed to be given by the Landauer formula 
for quantum point contact as 
\begin{equation}
R(T) = R_0 \cdot [ 1 + \exp(E_{\rm c}/T) ], 
\end{equation}
where $E_{\rm c}$ is the threshold energy for the electron transmission 
measured from the chemical potential. 
The junction is insulating for $E_{\rm c} > 0$ 
and metallic for $E_{\rm c} <0$. 
This model is appropriate for relatively thin films 
where contacts among islands are relatively weak. 
For such a weak contact 
the resistance exhibits poor metallic or insulating temperature dependence. 
In Fig. 1 the temperature dependence of the resistance 
of a type-N junction is shown for a poor metallic case. 
While the resistance saturates at low temperatures, 
it exhibits almost exponential temperature dependence 
in some temperature range. 

In the following we consider the circuit 
consisting of 3 conductors connected in parallel. 
Each conductor consists of 10 composites 
of resistors connected in series. 
Each composite consists of 3 resistors 
connected in parallel. 
Each resistor is modeled by 
either type-J or type-N junction. 
Although circuits in general cannot be reduced 
to series-parallel combinations of resistors, 
we adopt this model, 3-10-3 circuit, for simplicity 
to discuss qualitative aspects of the resistance of a random network. 

In Fig. 2 
the temperature dependence of the resistance 
for 3-10-3 circuit is shown in the case 
where every resistor in the circuit is type-J junction. 
The values of $R_{\rm N}$ for junctions are randomly chosen. 
The resistance vanishes faster than single exponential function of $T$ 
as $T$ is decreased. 
This temperature dependence 
is similar to that obtained in the former study.~\cite{Dynes} 

In Fig. 3 
the temperature dependence of the resistance 
for 3-10-3 circuit is shown in the case 
where every resistor in the circuit is type-N junction. 
The values of $E_{\rm c}$ are randomly chosen. 
A crossover from metallic to insulating behavior is seen 
as $E_{\rm c}$ is increased. 
This crossover 
is similar to that obtained in the former study.~\cite{Meir} 

In Fig. 4 
the temperature dependence of the resistance 
for 3-10-3 circuit is shown in the case 
where a resistor is either superconducting or metallic. 
Almost exponential temperature dependence is seen 
reflecting the behavior of metallic junctions. 
This behavior is consistent with experiments, 
while it was unexplained in the former study~\cite{Dynes} 
where only superconducting junctions were taken into accounts. 

In experiments 
a S-M-I crossover~\cite{crossover} is observed 
when the film thickness is tuned. 
Such a crossover is also seen in Figs. 2-4. 
In our phenomenological analysis 
it is reduced to the nature of each junction 
and has nothing to do with a macroscopic phase transition. 
It should be also noted that the present metallic state is 
irrelevant to the recent issue~\cite{MIT} 
of the presence of a metallic state 
in two-dimensional disorderd systems. 
The issue corresponds to the presence of a coherence 
at macroscopic scale in uniform system after averaging. 
On the other hand, 
a metallic state is possible at mesoscopic scale. 
Our system is a mesoscopic one in terms of junctions, 
while superconducting islands are macroscopic objects. 
The system is highly heterogeneous 
where the coherence is maintained within the islands 
and the loss of the coherence occurs only at the junctions. 

The author is grateful to 
T. Kawaguti, B. Shinozaki and K. Makise 
at Kyushu University Ropponmatsu 
for valuable discussions. 

\vskip 30pt

%----------------------------------------------------------------

%----------------------------------------------------------------
\vskip 15pt
%----------------------------------------------------------------
\begin{figure}
\caption{
Temperature dependence of the resistance 
for a single normal junction. 
The threshold energy is chosen as $E_{\rm c}/T_{\rm c}= -1.5$. 
}
\label{fig1}
\end{figure}

\vskip -15pt

\begin{figure}
\caption{
Temperature dependence of the resistance 
for a network of superconducting junctions. 
The normal resistances are randomly distributed in the range, 
$0.1 < R_{\rm N}/R_0 < 0.6$, for the bottom case. 
The range for the middle case is $0.2 < R_{\rm N}/R_0 < 0.7$ and 
$0.3 < R_{\rm N}/R_0 < 0.8$ for the top case. 
}
\label{fig2}
\end{figure}

\vskip -15pt

\begin{figure}
\caption{
Temperature dependence of the resistance 
for a network of normal junctions. 
The threshold energies are randomly distributed in the range, 
$-0.4 < E_{\rm c}/T_{\rm c} < 0.6$, for the top case. 
The range for the other cases is shifted as 
$-0.5 < E_{\rm c}/T_{\rm c} < 0.5$, 
$-0.6 < E_{\rm c}/T_{\rm c} < 0.4$, 
$-0.7 < E_{\rm c}/T_{\rm c} < 0.3$ and 
$-0.8 < E_{\rm c}/T_{\rm c} < 0.2$ in order (form top to bottom). 
}
\label{fig3}
\end{figure}

\vskip -15pt

\begin{figure}
\caption{
Temperature dependence of the resistance 
for a network of the random mixture of superconducting and normal junctions. 
The normal resistances for superconducting junctions 
are randomly distributed in the range 
$0.1 < R_{\rm N}/R_0 < 0.6$. 
The threshold energies for normal junctions 
are randomly distributed in the range 
$-1.5 < E_{\rm c}/T_{\rm c} < -1.0$. 
For the top case 3\% of the junctions are superconducting 
and 97\% are normal. 
For the middle case 5\% are superconducting. 
For the bottom case 7\% are superconducting. 
}
\label{fig4}
\end{figure}
%----------------------------------------------------------------
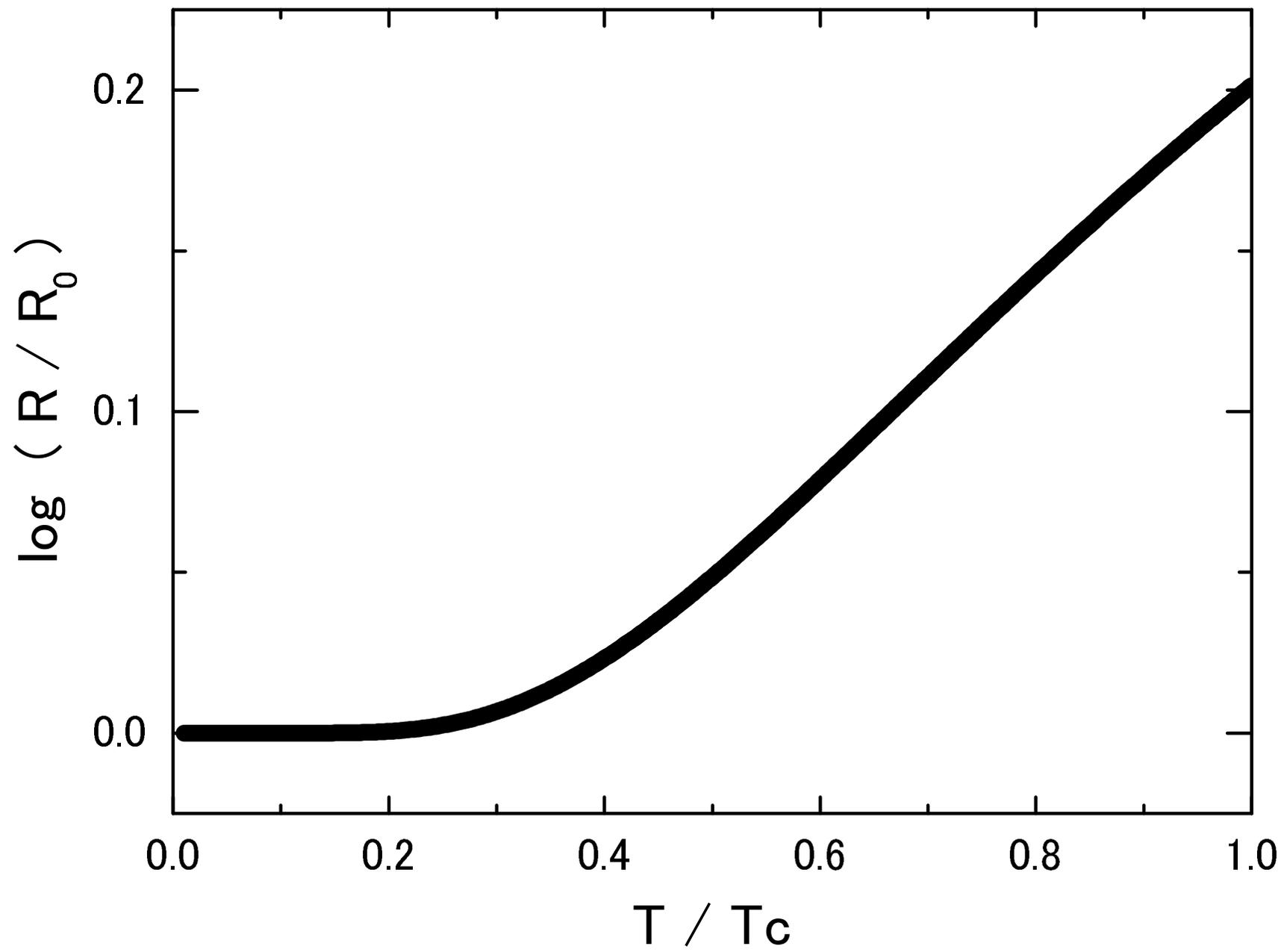
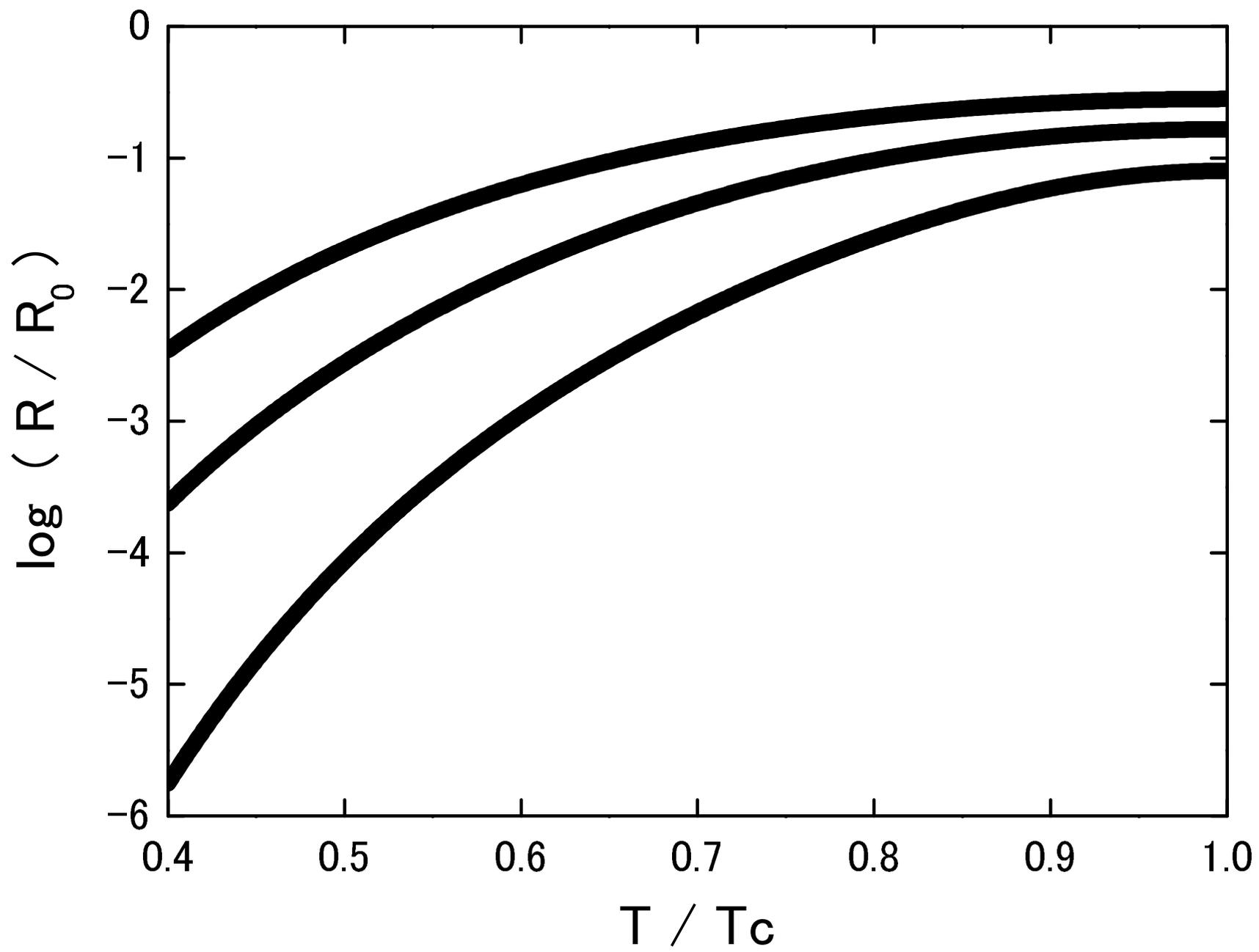
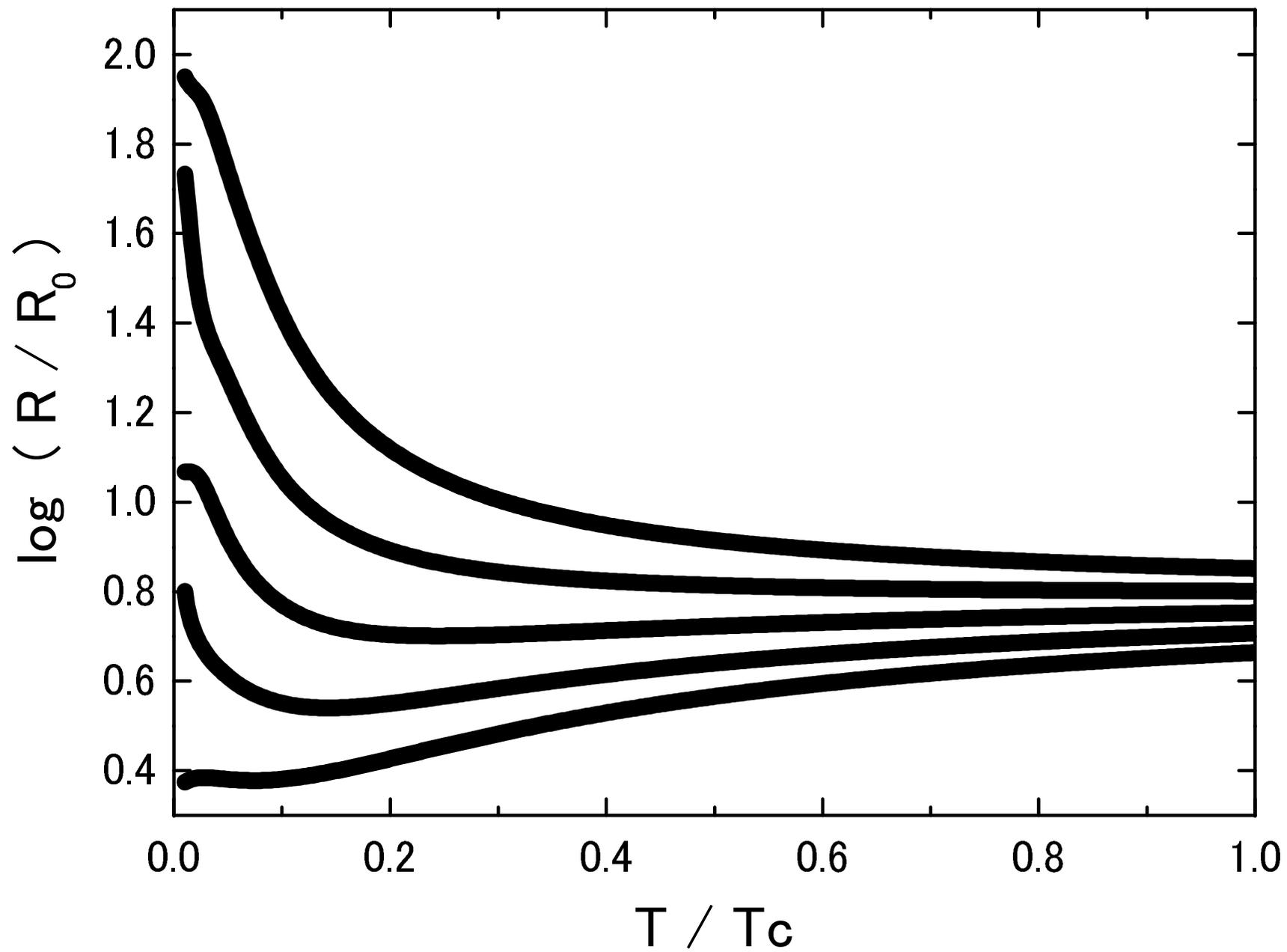
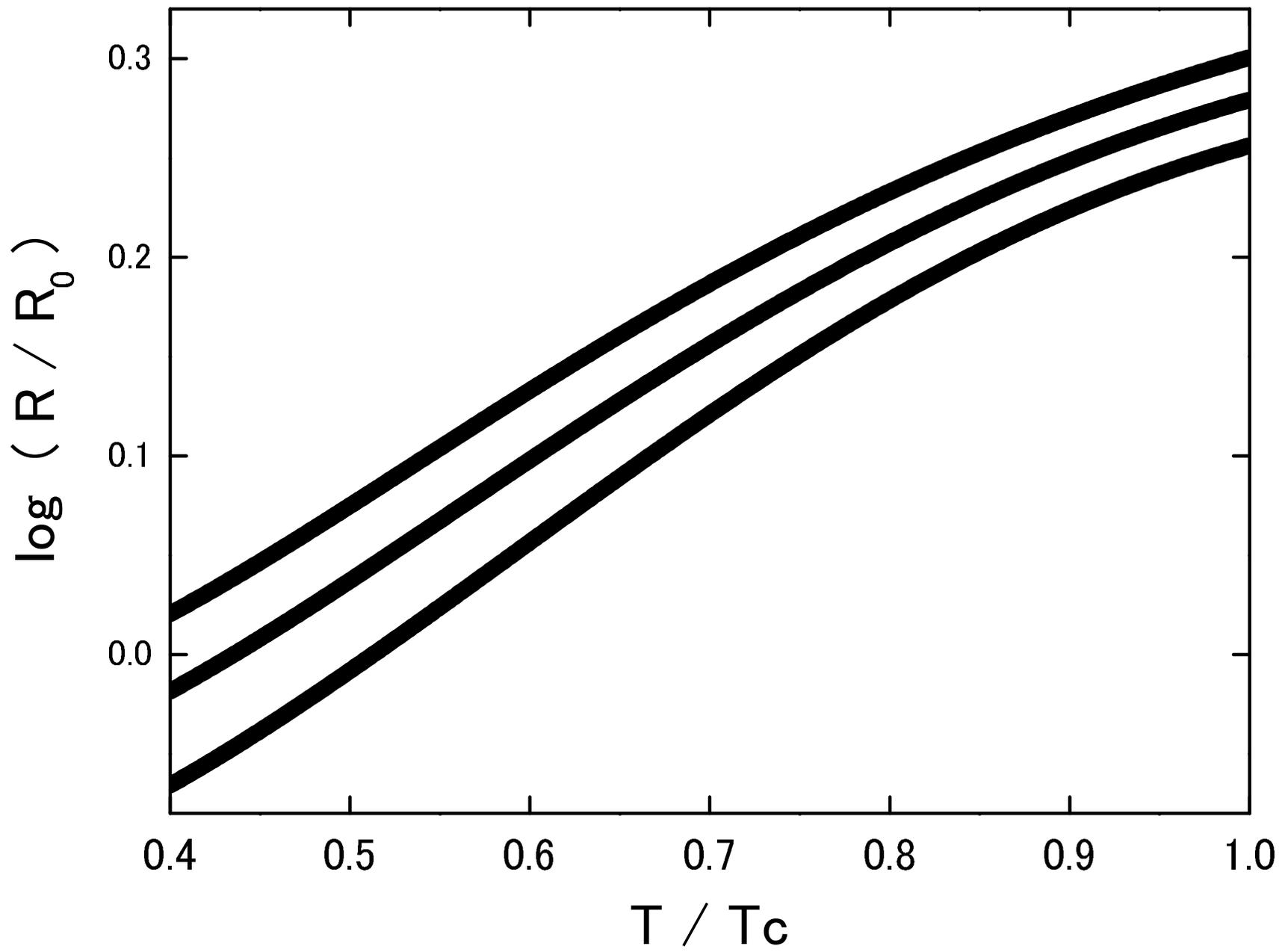
\end{document}